\documentstyle[11pt,aasms]{article}

\def\etal{{\it et al.\ }}
\def\pcc{\,{\rm cm^{-3}}}
\def\kel{\,{\rm K}}
\def\cmm{\,{\rm cm^{-2}}}
\def\kpc{\,{\rm kpc}}
\def\Mpc{\,{\rm Mpc}}
\def\yr{\,{\rm yr}}
\def\ltsima{$\; \buildrel < \over \sim \;$}
\def\lsim{\lower.5ex\hbox{\ltsima}}
\def\gtsima{$\; \buildrel > \over \sim \;$}
\def\gsim{\lower.5ex\hbox{\gtsima}}

\begin{document}

 \title{TWO--PHASE COOLING FLOWS WITH MAGNETIC RECONNECTION}

 \author{Colin Norman\altaffilmark{1}}

 \centerline{and}

 \author{Avery Meiksin\altaffilmark{2}}

 \affil{ Institute for Theoretical Physics, University of California,
Santa Barbara}

 \altaffiltext{1}{ Johns Hopkins University, Department of Physics and
Astronomy, 3400 N. Charles Str., Baltimore, MD\ 21218;\ norman@stsci.edu}

 \altaffiltext{2}{ University of Chicago, Department of Astronomy and
Astrophysics, 5640 S. Ellis Ave., Chicago, IL\ 60637;\ Edwin Hubble Research
Scientist;\ meiksin@oddjob.uchicago.edu}

 \slugcomment{To appear in The Astrophysical Journal (1 September 1996)}

 \begin{abstract} Motivated by the observations of high Faraday
 rotation measures measured in cooling flow clusters we propose a
 model relevant to plasmas with comparable thermal and magnetic pressures.
 Magnetic field reconnection
 may play a major role in changing the topology of the magnetic field
 in the central cooling flow regions. The effect of the topology
 change is that {\it cool} flux loops can reconnect to {\it hot} flux
 loops that are connected to the overall thermal reservoir of the
 cluster. There can be a rapid recycling of mass between hot and cold
 phases on a time scale of $\sim 3 \times 10^8-10^9~\yr$ which may
 reduce the inferred inflow and mass condensation rates by at least an
 order of magnitude. A central multiphase medium is a direct
 consequence of such a model. Throughout the cooling flow the filling
 factor of the hot loops ($T >2 \times 10^7 \kel$) is of order unity. The
 filling factor of the cool loops($ < 2 \times 10^7 \kel)$ is $0.1 - 1
 \%$ with a corresponding mass fraction of cold phase of $1-10\%$. A
 crucial parameter is the coherence length of the field relative to
 the cooling radius and the distribution of field energy with scale.
 When the cooling radius is greater than the field coherence length
 then cooling flows proceed as usual. When the coherence length is greater
 than the central cooling radius, however, the thermal energy
 of the reservoir can be tapped and the mass condensation rates may be
 very significantly reduced. Three additional conditions must be
 satisfied: I.~Cold loops must be able to fall at least as far as the
 mean distance between hot loops in a cooling time; II. Loops must
 enter an evaporative phase on reconnecting; and III. A sufficient
 number of hot loops penetrate the cold phase region to power the
 radiative losses.

\end{abstract}

 \section{INTRODUCTION}
 
 Calculation of the mass flow rate into the centers of clusters of
 galaxies with short cooling times gives large values of order $\sim 100
 -1000 {M_\odot}{\rm yr}^{-1}$ (Binney and Cowie 1981). This point has
 dominated most of the thinking on the thermal evolution and central
 gas dynamics of clusters of galaxies (Fabian 1994). We propose a
 mechanism that will reduce these mass inflow rates by an order of
 magnitude in clusters where the coherence length of the field is
 greater than the cooling radius. The mechanism we invoke here
 involves, however, the study of turbulent magnetized fluids. Given
 our state of knowledge of the properties of these flows
 our calculations can, at present, only be regarded as
 indicative. However, the stabilization does seem to be a reasonable
 mechanism and we present it here in the spirit of the pioneering
 feedback work of Tabor and Binney (1993) and Binney and Tabor (1995).

 Our motivation for rethinking this subject was the beautiful data
 from Ge (1991), Ge and Owens (1993), Owens et al. (1990) clearly
 showing that cooling flow clusters have rotation measures between
 10--100 times those that have no established cooling. The inferred
 field strengths are at or near equipartition values with the thermal
 pressure. Magnetized clusters have been studied before in detail by
 Soker and Sarazin (1990). They assumed that once the magnetic field
 built up it was annihilated down to equipartition values via
 reconnection process. Normally, however, reconnection does not
 annihilate magnetic fields (Taylor 1974). Even with fast
 reconnection, annihilation of significant volumes of magnetic energy
 density does not take place. To dissipate the field globally requires
 a global anomalous resistivity and this is very difficult both to
 achieve and sustain except in very small volumes.

 The issues we have to address are complex. Magnetic field
 reconnection normally rearranges the field and may allow some change
 in the mass-to-flux ratio of loops.  Since the magnitude of the
 field does not change significantly during reconnection the magnetic
 pressure is not significantly altered.  The crucial point of physics
 for this work, however, is that reconnection significantly alters the
 topology and connectedness of the magnetic field.

	 In the central regions of cooling flows the line widths are
 random (Heckman et al. 1989) and it is most likely that the fluid is
 turbulent. The most recent analysis of the nature of thermal
 conduction in cooling flows (Tao 1996) indicates that there may be
 some but not complete suppression of the thermal conductivity. Additional
 suppression may result from electromagnetic instabilities in the cluster
 gas (Pistinner \etal 1996).
 We will adopt the values consistent with Tao's work in our subsequent
 calculations although we regard the thermal conductivity in a tangled
 field as still an unsettled question (cf. Rosner and Tucker 1989,
 Tribble 1989, Chun and Rosner 1993).

 Recent spectral X-ray measurements have established that the X-ray
 emitting gas is composed of two (or more) phases (Fukazawa \etal 1994;
 Fabian \etal 1994). We incorporate this important new datum into our model.
 Our proposed solution to the suppression of the large mass drop out
 rates in cooling flows is to recycle the cooling ( $\leq 10^7 \kel$) gas
 back into the hot phase ($ \geq 10^7 \kel$) and then allow it to cool
 again. Only a small fraction of the total cooling mass need be in the
 cool phase to match the observed low-temperature X-ray emission. The
 mass condensation rate is, then, the amount of cool gas per recycling
 time. Since current estimates of the amount of condensed mass are
 calculated assuming that the cooling gas eventually leaves the flow with no
 recycling, the currently estimated masses can be suppressed by the
 ratio of the recycling time to the age of the cluster.

 More specifically, the recycling solution we propose is as follows:

 The magnetic field lines engulfed by the outward moving cooling wave
 have a spectrum of scales. The ones with scales greater than the
 scale of the cooling wave remain connected to the essentially
 infinite heat reservoir of the cluster. Even with the turbulent
 thermal conductivity given by Tao (1996), field lines
 topologically connected to the outside of the cooling wave region
 remain hot and can be treated as isothermal. These hot tubes are the
 pipes that the cooling flux loops can tap into. Smaller scale
 cooling loops are not connected to the heat bath of the
 cluster. However, small-scale cooling loops can be heated and recycled by the
 process of magnetic field reconnection giving the required recycling
 mechanism.

 In the next section~(2) we set up a simple physical model and
 present estimates of the characteristics of our solution.  Thermal
 structure and recycling are discussed in more detail in section~3.
 In Section~4 we discuss some observational consequences and a
 summary and conclusions are given.

 \section{BASIC MODEL AND ESTIMATES}

	       Assume an initially tangled isotropic field across the
 cluster with a plasma parameter $ \beta = (P_{\rm thermal} /
 P_{\rm magnetic}) = \beta_0 \geq 10$ since the cluster is initially
 dominated by thermal pressure and not magnetic pressure. We assume
 there is a spectrum of loops as a function of scale. We envisage these
 intracluster fields to be generated by radio galaxies, outflows from
 AGNs and starburst galaxies and the outflow processes from normal
 galaxies that enrich the intracluster medium that are presumably from
 a supernovae driven source. With these processes in mind we estimate
 that the scales of the fields in the general intracluster medium can
 range from $ \sim ~10~\kpc$ up to $\sim ~1~\Mpc$. The field may be
 completely coherent across the cluster if there is a large scale
 primordial field or if large scale fields are generated in the process
 of structure formation (Kulsrud 1995). 

	       After a time $t$ the material enveloped by the
 outwardly propagating cooling wave (with radius defined by that
 radius at which $ t_{\rm cool} = t$) can cool and move inward
 (Bertschinger 1989).  In the kinematic inflow phase the fields are
 frozen in and the stretching and amplification that occur are well
 described in the paper by Soker and Sarazin (1990). The field becomes
 increasingly radial and we will assume for simplicity of analysis
 that it is exactly radial in this phase.

 Assuming complete ionization, the cooling law that describes the
 propagation of the cooling wave is given by:
$$
  {3 \over 2} {dnkT \over dt} = n^2 \Lambda(T), \eqno(1)
$$
 where $T$ is the gas temperature, $n$ is the density, and
 $\Lambda(T)$ is the cooling function. 
 We assume that the gas and dark matter both have the distribution of
 an isothermal sphere with a gas temperature, $T_g$, and dark matter
 temperature, $T_d$, with gas density
$$
 n = n_0 ({ r_0 \over r})^{\alpha} . \eqno(2)
$$
where $\alpha =  {2 T_d / T_g}$ and the subscript 0 refers to gas properties
at a fiducial radius $r_0$. The outwardly propagating cooling wave has the
solution:
$$
 r_{\rm cool} = r_0 \left( t \over t_{\rm cool,0}\right)^{1/\alpha}, \eqno(3)
$$
with 
$$
  v_{\rm cool} = v_{\rm cool,0}\left( {r_0 \over r_{\rm cool}}\right)^
{ \alpha - 1}, \eqno(4)
$$
where 
$$
  t_{\rm cool,0}= \left[{ 3 kT_0 \over 2 n_0 \Lambda(T_0)}\right], \eqno(5)
$$
and $v_{\rm cool,0}= (r_0 / t_{\rm cool,0})$. The mass flux into the cooling
bubble, at any given time $t$ and at radius $r_{\rm cool}$, is:
$$
  {\dot M_{\rm cool}} = {\dot M_0} \left({r_{\rm cool}\over r_0}\right)^
{3-2\alpha}, \eqno(6)
$$
where 
$$
 {\dot M_0} = \left( {4 \pi \rho_0 r_0^3 \over t_{\rm cool,0}}\right).
\eqno(7)
$$
As the cooling wave propagates outward, the material trapped in small
magnetic loops will start to cool and fall inward.

 The plasma parameter $\beta$ varies as a function of radius as:
$$
  \beta = \beta_0 
 \left( {r \over r_0}\right)^{ \alpha/3}, \eqno(8)
$$
 and the flow can be significantly magnetized when $\beta \sim 1$. We
 are assuming that the flow is sufficiently large as to entrain
 field lines into the central regions of the cooling flow, allowing a
 build up of the magnetic pressure. The increased magnetic pressure does
 {\it not} change the estimates of the accretion rates that are based
 purely on the thermal properties, not the momentum balance.

 As discussed by Bertschinger and Meiksin (1986) and Meiksin (1988), if the
 thermal conductivity is within an order of magnitude of the Spitzer
 value, the cooling along flux tubes that are simply connected to the
 outside of the cooling wave is negligible. Therefore these connected
 flux tubes do not contain any cool material. The flux tubes that are
 not connected can cool. This leads to a two-phase medium of hot
 connected flux tubes and cool disconnected tubes. 
 As discussed by Soker and Sarazin (1990) even if all the flux were
 reconnected and dissipated the associated heat input would have
 negligible effect on the thermal balance.

 Reconnection is a natural way to recycle the cold material back into
 the hot phase. The exact details are unclear but the result is
 probably bounded by the rules for collisions of flux tubes
 (cf.\ Norman \etal 1996) as a lower bound or
 explosive reconnection with all the mass injected into the background
 medium between hot and cold phases as an upper bound.

 Let us assume that a fraction, $f_h$, of the mass and flux is in hot
 tubes and the rest in cold tubes. Let us also assume that there is a
 mean flux reconnection rate, $\Gamma$, giving a mean
 mass exchange rate ${\dot M_e}$ between hot and cold of:
$$
   {\dot M_e} = \left( {\Phi \over M}\right)^{-1} \Gamma \Phi, \eqno(10)
$$
 where the flux to mass ratio must be taken in mean averaged sense
 here. Therefore since the timescale for mass exchange is $\Gamma^{-1}$,
 the fraction of the total mass, $M_h$, within a cooling radius that is in
 the form of cool condenstates ( $ < 10^7 K$) is:
$$
  M_{rec} = \left(\Gamma \tau\right)^{-1} M_h, \eqno(11)
$$
 where $\tau$ is the time for mass to leave the hot phase, which is on
 the order of the age of the cluster because of thermal heat conduction. The
 amount of mass being recycled at any given time is negligible compared to the
 total amount of mass in the hot phase.

 Concentrating on one loop and using the fast reconnection law, Soker
and Sarazin (1990) give a typical reconnection timescale,
$\tau_R=\Gamma^{-1}\sim l_t/(\epsilon V_A)$ of:
$$
\tau_R = 10^8 \yr \left({\epsilon \over 0.1}\right)^{-1}\left({ l_t
\over 3 \kpc}\right)\left( { B \over {100 \mu G}}\right)^{-1}\left( {n
\over 0.03 \pcc}\right)^{ 1\over 2}. \eqno(12)
$$
 for our parameters. We assume that reconnection occurs with a
 reconnection velocity of $ \epsilon$ times the Alfven velocity. The
 characteristic scale length for reconnection is here taken to be the
 transverse dimesion of a loop. Note that when large scale hot loops
 reconnect with small-scale cool loops the reconnection time given in
 (12) should be regarded as an upper limit since the reconnection time
 may be considerably shorter due to the smaller size of the loops. The
 internal Alfven speed $V_A$ in the cool and hot loops does not change
 significantly since $V_A \propto n^{\gamma}$ where $0 \leq \gamma \leq
 1/6$. Complete reconnection will occur on a single collision if the
 reconnection time is less than or equal to the crossing time of the
 loops. If this is not satisfied then collisions with hot loops will
 continue until the process is complete. This is similar to
 shredding. For more details regarding incomplete reconnection during
 flux loop collisions see Norman \etal (1996). Generally, the complex
 process of shredding of loops and blobs as they fall into the central
 regions of the cooling flow is one where cool gas distribution
 cascades to smaller and smaller scales. A small inefficiency in the
 recycling would permit some material to continue to cool and make
 stars, but at a much reduced rate compared to the standard cooling
 flow prediction.

 We note here that buoyancy is an inefficient means for transporting flux out
 of the cooling region (Soker and Sarazin 1990). Flux loops will be
 buoyant and reach an appropriate scale height. We assume that most
 of the loops are still trapped, although there is probably still some
 flux loss. Reconnection will then occur and the value of mass and flux
 for the product loops will change. There will then be a rising and
 falling of loops as the reconnection process continues and re-sorts
 the mass and flux, although it is not possible to follow an individual
 piece of flux and mass. Magnetic tension in reconnected loops may lift
 the cool optically-emitting material out of the center
 of a cooling flow, accounting for the wide range of scales over which the
 optical filaments are observed (Zoabi \etal 1996).
 In practice, loops will shred into smaller
 structures due to Kelvin-Helmholtz instabilities after only a few
 Brunt-V\"as\"ail\"a oscillation periods (Reale et al 1991). As
 discussed below this will enhance their evaporation and mass
 recycling.

 \section{THERMAL STRUCTURE AND RECYCLING}

 We now describe the thermal structure of the intracluster gas. The
 radiative cooling of the hot phase is nearly balanced by heat conduction
 from the outer portions of the cluster into the cooling region by
 those extended flux tubes that connect the external heat reservoir to the
 cooling interior. While perfect balance between cooling and heat conduction
 requires too high an external temperature compared to observations, acceptable
 solutions are found when a small amount of inflow is allowed (Bertschinger
 \& Meiksin 1986), provided the temperature decline is not more than by a
 factor of $\sim2-3$ (Meiksin 1988). Accordingly, we may expand the fluid
 variables in the small parameter $\epsilon=\dot M_{\rm flow}/\dot M_h$, where
 $\dot M_{\rm flow}$ is the true accretion rate, and $\dot M_h$, given by
 equation (7) above and evaluated at the cooling radius, is the rate that would
 be inferred from the X--ray luminosity assuming heat conduction were
 suppressed. Thus, for example, we may write, to first order, $n(r)=n_0
 (r/r_{\rm cool})^{-\alpha}[1+\epsilon n_1(r)]$ and $T_h(r)=T_0(r)
 [1+\epsilon T_1(r)]$. The zeroth order equation for pressure support (both
 thermal and magnetic) then requires, for $\beta>1$,
$$
\alpha = {2T_d \over T_h}\left[1+\beta(r)^{-1}\right]^{-1}, \eqno(13a)
$$
when the gas temperature varies slowly with radius compared to the density.
 We point out that in the case $T_d \sim T_h$ and where the
 cooling flow is sufficiently magnetized with $\beta \sim 1$, then
 $\alpha \sim 1$. Consequently, from equation (6), $\dot M_{\rm cool}
 \propto r $, a case favored by the observations. We will regard $\alpha$
 below as being directly determined from the measurements of the X--ray
 surface brightness of the hot component, since the radial dependence of the
 surface brightness depends only weakly on that of the temperature.

 Setting $x=r/ r_{\rm cool}$ and $y=T_0(r)/ T_h$, the zeroth order
 thermal energy equation becomes
$$
{2\over7}{1\over x^2}{d\over{dx}}\left(x^2{{dy^{7/2}}\over{dx}}\right)=
\left({r_{\rm cool}\over\Lambda_{\rm F, c}}\right)^2x^{-2\alpha},
\eqno(13b)
$$
where $\Lambda_{\rm F, c}=[\kappa(T_h)T_h/ n_0^2\Lambda)]^
{1/2}$ is the Field length, evaluated at the cooling radius (Begelman and McKee
1990; Field 1965). Near $T\sim2\times10^7\, {\rm K}$, the cooling function
is nearly independent of $T$. The Field length is then
$$
\Lambda_{\rm F,c}\approx100\kpc \left({{0.003\pcc}\over n}\right)
\left({T_h\over{2\times10^7\kel}}\right)^{7/4}, \eqno(14)
$$
where we have used the cooling function of McKee and Cowie (1977). The
solution of equation (13b) is then
$$
T(r) = T_h\left[{7\over4}\left({{r_{\rm cool}}\over{\Lambda_{\rm F,c}}}
\right)^2{1\over{(2\alpha-3)(\alpha-1)}}\left({1\over x^{2\alpha-2}}
-{1\over x}\right)+{1\over x} + y_\infty^{7/2}\left(1-{1\over x}\right)
\right]^{2/7}, \eqno(15a)
$$
for $\alpha\neq3/2$ and $\alpha\neq1$;
$$
T(r) = T_h\left[-{7\over2}\left({{r_{\rm cool}}\over{\Lambda_{\rm F,c}}}
\right)^2{{\log x}\over x}+{1\over x} + y_\infty^{7/2}\left(1-{1\over x}\right)
\right]^{2/7}, \eqno(15b)
$$
for $\alpha=3/2$, and
$$
T(r) = T_h\left[{7\over2}\left({{r_{\rm cool}}\over{\Lambda_{\rm F,c}}}
\right)^2\log x + {1\over x} + y_\infty^{7/2}\left(1-{1\over x}\right)
\right]^{2/7}, \eqno(15c)
$$
for $\alpha=1$. Here, $y_\infty$ is an integration constant.

 The solution given in equation (15) is a gradually decreasing temperature
 profile toward the center. A typical cooling flow will have a cooling radius
 of $\sim100\kpc$ and $T_h\sim5\times10^7\kel$, giving $r_{\rm cool}/
 \Lambda_{\rm F,c}\sim0.3$. For the simple, observationally favored case
 of $\alpha \approx1$ and $y_\infty\approx1$, we find that the temperature at
 $r\sim0.1r_{\rm cool}$ will be below its value at the cooling radius by a
 factor of $\sim(1-0.3\log 10)^{2/7}\approx0.7$, consistent with {\it ASCA}
 measurements for the hot phase (Fukazawa \etal 1994).

 A strong argument in favor of high mass drop-out rates is the high
 luminosities measured in the low-temperature X-ray emission lines,
 such as the Fe L lines (Mushotzky \etal 1981; Fabian \etal 1994).
 These lines demonstrate the presence of low temperature
 gas that is cooling at a high rate, and buttress the argument for a
 correspondingly high rate of mass condensation out of the flow. We next
 show how recycling may significantly alter this point of view.

 Consider two phases in pressure equilibrium, a hot phase and a cold
 phase at respective temperatures $T_h$ and $T_c$, with number
 densities $n_h$ and $n_c$.  The cold phase arises from thermally
 unstable gas suspended in magnetic flux loops. Let $f_c$ denote the
 volume filling factor of the cold phase. The mass $M_c$ in cold phase
 is related to that in the hot, $M_h$, by

 $$M_c = {f_c\over{1-f_c}}{n_c \over n_h}M_h \approx f_c{T_h\over T_c}M_h.
\eqno(16)$$

 \noindent The mass cooling rate $\dot M_c$ of the gas in the cold phase is
 then related to that in the hot, $\dot M_h$, by

 $${\dot M_c} = f_c\left({T_h\over T_c}\right)^3{{\Lambda(T_c)}\over
 {\Lambda(T_h)}}{\dot M_h}, \eqno(17)$$

 \noindent for a cooling function $\Lambda(T)$.
 In principle, it is possible for the cooling rate inferred from emission lines
 generated by the cold phase to exceed that inferred from the hot. We will
 show below, however, that in our model these rates should be comparable,
 though some variance in the ratio is to be expected.

 The crucial point now is that as the flux loops carrying the cold
 phase sink they will reconnect to the hot flux tubes. Thermal heat
 conduction carried along the hot flux tubes will then quickly reheat
 the cold phase gas to $T_h$, resulting in a continual recycling of
 the cold phase material. The actual mass drop-out rate will thus be
 much less than inferred from the line luminosities. This is possible
 provided three criteria are satisfied:

 I.~The cold loops are able to fall at least as far as the mean
 distance $\lambda_h$ between hot flux tubes in a cooling time:
 $v_{ff}t_{\rm cool,c}>\lambda_h$;

 II: Loops must enter an evaporative phase on reconnecting;

 III.~There is a sufficient number of hot flux tubes penetrating the cool phase
 region to conduct heat inward from the infinite heat bath surrounding the
 cool phase to power the total radiative losses from both the cool and hot
 phases.

These conditions may be met as follows:

 I.~To make a simple estimate assume that for the purposes of argument
that the cooling time is of the order of the time for a loop to cross the
cooling region. Then for a {\it volume} filling factor of the hot phase
within the cooling region of $f_h \sim 1$, the covering factor of the
hot phase is $ \Omega_h \sim f_h^{2 \over 3} \sim 1$.  The cool loops
will then collide with the hot phase on the order of a cooling time of
the cool phase.

II. The cool loops must have a characteristic scale less than the
Field length $\Lambda_F$ when they reconnect to the hot
loops.  For loops that are initially magnetically isolated and collide
early on ( when they have temperatures of order, say, $3 \times 10^6 \kel$)
with hot loops they will undergo evaporation in the classical
limit and be recycled since
$$
\Lambda_F = 3 \kpc \left({ 0.03 \pcc \over n} \right)\left({ T_h \over
3 \times 10^7 \kel} \right)^{5/2} \left({ T_c \over 3 \times 10^6 \kel}
\right)^{1.6}, \eqno(18)
$$
where we have allowed for the radiation of the cooler material using the
cooling approximation given in McKee and Cowie (1977).

Evidence for absorption by very cold clouds has been argued for on the basis
of the X-ray spectra in some cooling flows, with inferred characteristic
cloud sizes of $\lsim 1$ pc (Fabian 1994). In our model, such clouds will
ultimately result from any small inefficiency in reheating. We suggest that
these clouds are shortlived, and will evaporate when they reconnect with
hotter loops. For the small cloud sizes suggested, heat conduction occurs
in the saturated limit, so that the clouds will disappear in an evaporative
wind (McKee and Cowie 1977). The timescale for the clouds to evaporate is
$$
t_{ev} = 10^8 \yr \left({10 K \over T_c }\right) \left( {T_h \over 3 \times
10^7 \kel}\right)^{ 1/2} \phi_s \left({ \sigma_0 \over 10^3}
\right)^{ 3/8}\left({ R \over 1 {\rm pc}}\right), \eqno(19)
$$
where $R$ is the cloud radius, and the saturation parameter, $
\sigma_0 = 4 (T_h/3 \times 10^7\kel)^2 /(n_h R_{\rm pc} \phi_s)$, is $\sim
10^3$, taking $\phi_s \sim 0.3$ (Balbus and McKee 1982).
   
 From analyses of absorption-line studies in both X-ray and radio bands
 the temperature of the clouds must be less than $10\kel$ (Crawford and
 Fabian 1992; Dwarakanath \etal 1994). The X-ray absorption measurements
 indicate \ion{H}{1} column densities of the order of $10^{20} - 10^{21} \cmm$
 (Fabian 1994). At the canonical pressure
 of the central cooling flow this corresponds to scales of only
 $10^{16}-10^{17}$ cm corresponding to cloud evaporation timescales of
 $\sim10^5 - 10^6 \yr$. Therefore, when the reconnection occurs the cold
 \ion{H}{1} will evaporate and be recycled into the hot phase.
 The cloud destruction time is given by the time for reconnection with
 hot loops ($\sim 10^8 \yr$). For unit covering factor the cloud
 destruction rate in the central region of order $ R_{\rm central} \sim
 1\kpc$ is $\sim 1 (R_{\rm central}/1 \kpc)^2 (N_{\rm HI}/10^{21} \cmm)
 M_{\odot} \yr^{-1}$.

 III.~The hot phase will occupy most of the volume.  A reasonable estimate
of the cooling filling factor is $f_c\approx(\Gamma \tau)^{-1} (T_c/T_h) \sim
10^{-3}<< 1 $. From equation (16), only about 1\% of the total intracluster
gas within the cooling radius will then be in the cool component. Because
of the relatively greater efficiency for generating X-ray photons per unit
mass of the cool phase compared to the hot, however, equation (17) shows
that the mass cooling rate of the cool component will be comparable to that of
the hot, and may even somewhat exceed it. If the cool phase were detected
through an X-ray emission line, such a disagreement, which could be interpreted
in the standard cooling flow picture as an abundance anomaly of the element
producing the line, could equally well be interpreted as evidence for a second,
denser and cooler phase.

\section{SUMMARY, CONCLUSIONS, AND OBSERVATIONAL CONSEQUENCES}

We have proposed a model for significantly reducing the mass inflow
rates in clusters of galaxies. Reconnection processes acting on the
topology of the field regulates the inflow of heat from the vast
heat bath of the entire cluster. Reconnection has no energetic
advantage at all; it is simply the topology change that allows
heating of the matter in cold loops and a relatively fast recycling
time. Therefore the cool gas can be used many times, in contrast to
conventional cooling flow models in which once the gas cools, it
necessarily becomes invisible and can never be recycled.  A crucial
parameter is the coherence length of the field relative to the cooling
radius and the distribution of field energy with scale.  When the
cooling radius is greater than the field coherence length, cooling
flows proceed as usual. When the coherence length is greater than the
central cooling radius, however, the thermal energy of the reservoir
can be tapped and the mass drop out rates can be very significantly
reduced. Three additional conditions must be satisfied: I.~Cold loops
must be able to fall at least as far as the mean distance between hot
loops in a cooling time; II. ~Loops must enter an evaporative phase on
reconnecting; and III.~A sufficient number of hot loops must penetrate
the cold phase region to power the radiative losses.

Central regions will be multi-phase with a hot-phase filling factor of
order unity and the cold-phase filling factor very much less than
unity. There is rapid mass circulation on a timescale of order $10^8$~yr.

Filaments will be highly magnetized. In general, the central regions
of cooling flows should look very magnetized and filamentary. Detailed
rotation measure maps could map the central field geometry. The
two-phase model we outline here is consistent with the recent {\it ASCA}
observations of the centers of cooling flows (Fabian et al. 1994)
although many details remain to be understood. 

We thank our colleagues for many useful and enjoyable discussions on
this subject in particular F.~Adams. S.~Cowley, M.~Donahue, T. Heckman,
D.~Melrose, R.~Mushotsky, R.~Sudan and M.~Voit. We thank the Institute for
Theoretical Physics at UCSB for the stimulating environment where this
interdisciplinary effort was possible. The ITP is funded by NSF Grant
No.\ PHY94-07194. A.M. is grateful to the W. Gaertner fund at the University
of Chicago for support.


\begin{references}

Balbus, S. A., \& McKee, C.F. 1982 \apj, 252, 529

Begelman, M.~C., \& McKee, C.~F. 1990 \apj, 358, 375

Bertshinger, E. 1989 \apj, 340, 666

Bertschinger, E.,\ \& Meiksin, A. 1986, \apj, 306, L1

Binney, J., \& Cowie, L. 1981, \apj, 247, 114

Binney, J.,\ \& Tabor, G. 1995, \mnras, in press

Chun, E.,\ \& Rosner, R. 1993, \apj, 408, 678

Crawford, C., \& Fabian, A.C. 1992 \mnras, 259, 265

Dwarakanath, K.~S., van Gorkom, J.~H., \& Owen, F.~N. 1994 \apj, 432, 469

Fabian, A. 1994, \araa, 32, 277

Fabian, A.~C., Arnaud, K.~A., Bautz, M.~W., \& Tawara, Y. 1994, \apj, 436, L63

Field, G.B. 1965 \apj, 142, 531

Fukazawa, Y. et al. 1994, PASJ, 46, L55

Ge, J.-P. 1991, PhD Thesis, New Mexico Institute of Mining and Technology

Ge, J.-P., \& Owens, F.~N. 1993, \aj, 105, 778

Heckman, T., Baum, S.~A., van Breugel, W.~J., \& McCarthy,
P.~J. 1989, \apj, 338, 48

Kulsrud, R.~M. 1995, Seminar ITP

McKee, C.F., \& Cowie, L.L. 1977, ApJ, 215, 213

Meiksin, A. 1988, \apj, 334, 59

Mushotzky, R.~F., Holt, S.~S., Smith, B.~W., Boldt, E., \& Serlemitsos, P.~J.
1981, \apj, 244, L47

Norman, C., Adams, F., Cowley, S., \& Sudan, R.~N. 1996, \apj, submitted

Pistinner, S., Levinson, A., \& Eichler, D. 1996 (preprint)

Owens, F.~N., Eilek, J.~A., \& Keel, W.~C. 1990, \apj, 362, 449

Reale, F., Rosner, R., Malagoli, A., Peres, G., \& Serio, S. 1991 \mnras, 251,
379

Rosner, R.,\ \& Tucker, W.~H. 1989, \apj, 338, 761

Soker, N.,\ \& Sarazin, C.~L. 1990, \apj, 348, 73

Tabor, G.,\ \& Binney, J. 1993, \mnras, 263, 323

Tao, L. 1996, \mnras, in press

Taylor, J. B. 1974, Phys. Rev. Lett., 33, 1139

Tribble, P.~C. 1989, \mnras, 238, 1247

Zoabi, E., Soker, N., \& Regev, O. 1996, \apj\ (in press)
\end{references}
\end{document}